**Title: Semiconductor nanostructures engineering: *Pyramidal* quantum dots**

**Authors:** E. Pelucchi, V. Dimastrodonato, L.O. Mereni, G. Juska and A. Gocalinska

**Affiliation:** *Epitaxy and Physics of Nanostructures, Tyndall National Institute, University College Cork, Cork, Ireland*

**Abstract**

Pyramidal quantum dots (QDs) grown in inverted recesses have demonstrated over the years an extraordinary uniformity, high spectral purity and strong design versatility. We discuss recent results, also in view of the Stranski-Krastanow competition and give evidence for strong perspectives in quantum information applications for this system. We examine the possibility of generating entangled and indistinguishable photons, together with the need for the implementation of a, regrettably still missing, strategy for electrical control.

**Introduction**

One of the open challenges for the epitaxial semiconductor quantum dot (QD) community -in all its branches and siblings: e.g. lasers and nanolasers, cavity quantum electrodynamics (QED), quantum information- is the improvement of the level of control on the single nanostructure (and/or the ensemble).

The idea of "*improving control*" is obviously a generic statement, which needs to be tailored and specified depending on the application and research field. A traditional QD laser will require a high density ensemble of QDs, as uniform in their emission properties as possible, to really exploit the concept of zero dimensional density of states. On the opposite side of QED, the need is that of precisely controlling the position, the spectral response and generally all quantum properties of a "single" artificial atom. In this context, the QD field uniformity becomes compelling only when scalability for "production" is needed, but can obviously be less of an issue if demonstrators of a particular quantum effect are the objective.

Different QD systems and approaches to diverse sets of controllable variables have been pursued in the scientific community, and some important breakthroughs have been obtained in the last few years as we will discuss in this contribution. Generally speaking a single QD system should ideally show most if not all of these characteristics: a) high optical quality (i.e. low or absent spectral meandering of the excitonic features), b) precise control over positioning (e.g. inside a photonic cavity or waveguide) and QD density, c) high uniformity as a guarantee of reproducibility and technological scalability, d) wavelength tunability (e.g. for matching atomic transitions), e) tailored symmetry properties (e.g. better than $C_{2v}$ for entangled photon emission), f) customized excitonic properties (i.e. tailored binding energies and dot population) and oscillator strengths, g) electrical "control" and many more.

In this review we will show that the "*Pyramidal*" quantum dot family (PQD, firstly developed successfully in EPFL by the group of Prof. Kapon)[1,2] is a valid candidate for QD engineering: it holds solid perspectives for quantum information technologies, having a high potentiality to meet most of the criteria just listed, and it turns out to be a nearly ideal system for exploring quantum dot physics and applications. We will often compare in this contribution PQDs to Stranski-Krastanow (SK) approaches (either modified to give site-control or simply self-assembled) so to establish a solid base for the discussion, addressing present challenges and giving some outlooks of expected attractive developments.

**State-of-the-art**

On the wish list of QD properties, control of the dot position certainly has to be one of the most addressed, and is definitely the one which has driven the scientific community the most in developing alternative approaches to the conventional SK approach (which, it must be said, despite the random nucleation and relatively poor uniformity has been the driving system for most if not all major recent breakthroughs in the "quantum" field, remaining the reference system for most researchers). It is in many senses the *"can't do without"* starting point of other controllable properties.

In this regard, the *Pyramidal* QD system is naturally a site-controlled system (for a review of early results see for example Ref. 3). The dots grow pseudomorphically to the (111) B substrate by metalorganic vapour phase epitaxy (MOVPE) thanks to anisotropic decomposition (from which growth rate anisotropies result) and capillarity effects [4] at the centre of an inverted tetrahedral pyramid (see Fig. 1), etched by standard or e-beam lithography into the GaAs substrate, exposing three (111)A facets.[5] No mask is present on the substrate during growth. As a consequence of the lithographic process, the QD location can be controlled on the substrate with nm-range precision, and their separation both in plane and along the growth direction can be handled at will. The dot shape is determined simply by the (self limited) profile of the underneath layers, their thickness and the composition (along with a delicate interplay of growth conditions). It must be said that a complex ensemble of quantum structures [three lateral quantum wires and a vertical one (VQWR), three lateral quantum wells (LQWs) and three vertical ones; the dot related structures solely are sketched in the inset of Fig.1] can be formed due to capillarity and capillarity induced segregation effects.[6]

It is worth mentioning that *capillarity* (i.e. fast diffusing atoms "tend" to fill the hole and flatten the profile, causing the QD thickening at the centre of the recesses which should reduce the total free surface and associated surface energy when compared to a pointed profile, as observable in Fig. 1) from the very beginning has been indicated as one of the effects allowing the dot formation (all this starting with V-groove quantum wires[7]). Nevertheless it was only recently that the source of growth rate anisotropies was undoubtedly clarified: i.e. the fact that the (Al)GaAs layers grow preferentially on the (111)A sidewalls, and significantly less on the planar unetched (111)B. This is the basis of the sharp profile which develops during growth, forming the template for the QD, *capillarity* induced, growth.[3,4] We will see that this has a strong consequence in terms of the uniformity in the broadest sense of the growth process. Here we will limit ourselves to underline that the growth process involves a preferential decomposition of the group III MOVPE growth precursors (e.g. Trimethylgallium) on the lateral (111) A facets, resulting in an anisotropic deposition rate of adatoms: i.e. adatoms (Ga, Al, In, etc) are preferentially deposited on the sidewalls, and only subsequently diffuse to form the crystalline layers.[4] As a consequence the sidewalls growth rate is higher, constantly maintaining a sharp profile at the bottom of the recesses (see Fig. 2a).[4] Fast diffusing adatoms (e.g. Ga, In) will try preferentially to fill the gap, generating the VQWR (e.g. a Ga rich region) at the centre if grown as part of an alloy (e.g. AlGaAs) and a QD when deposited alone.

In this last case the quantum confinement is given mostly by the barrier material (e.g. AlGaAs), but also by the fact that the other nanostructures attached to the dot (i.e. lateral quantum wires and quantum wells, see inset Fig. 1) are well thinner than the dot itself, and their quantum states are either not confined or lying at higher energies than those of the dot, so actually confining the QD.

The PQD growth process is in fact purely epitaxial ("orderly upon", etymologically), i.e. is a pure layer by layer process, like for 2D quantum wells. Differently from QWs, though, in this case there is the additional phenomenon of the thickening of the layers at the centre of the recesses. Everything, nevertheless, happens strictly pseudomorphically, with no relaxation processes, or change in growth mode like in SK dots.

On the contrary, in the SK case a number of techniques have to be applied to force the dot nucleation in a particular position. For clarity, hereunder we quickly discuss two of the most successful to date, as we will be comparing to these especially in this manuscript (neglecting the many possible variations on the theme[8,9,10,11,12]).

*The site-controlled SK in a hole approach (SKH).* The basic idea that naturally came to researcher's mind to control the dot position has been basically that of forcing the SK nucleation process inside a patterned recess, avoiding the nucleation process outside the recesses (typically shallow holes at the GaAs surface). It's worth mentioning that at first the idea faced several difficulties linked to the lithographic steps leaving behind either dirty or defected regions, and for a few years it had not been possible to see reports of good site controlled QDs by this approach. After some time of struggling, the community at last managed to develop processes without associated defects and finally good optical properties from ensemble and single site controlled QDs were obtained (we will discuss on this point more in detail later in the manuscript).[13,14]

Some growth issues associated with this approach are exquisitely linked to the competition between the nucleation in the patterned seed and on the planar region and/or at the holes (and/or their boundaries). The result is unwanted QD formation either on the planar regions or at the boundaries of the patterned sites. These problems have been addressed and mostly resolved in these years, but sometimes the solutions force specific epitaxial recipes, partly at cost of the versatility of the system. Moreover it should be said that this approach with SK dots (still InAs based) on InP started somehow later than on GaAs.[15] It is also worth mentioning that "capillarity" has necessarily a role also in these nucleation processes, and will act in competition to strain induced effects: on one hand complicating the description of the nucleation process, and on the other reducing the conceptual differences with PQDs.

*The SK on a ridge/pyramid (SKR) approach* [known as selective area grown (SAG) dots] has been developed as an alternative to the direct growth of SK dots on a patterned surface, to avoid the crystal quality inconveniences many groups encountered. The pattern is indeed present on the sample surface (as a $SiO_2$ mask for example on an (100) GaAs or InP substrate), but it is there only to select the wafer surface available for growth. A relatively thick growth is performed (somehow similarly to the PQDs), resulting in an out-rising ridge 2D structure, or in a 3D Pyramid.[16] During growth the top planar surface available can be reduced to a level which allows the growth of single SK dots on top of the structure.[17,18] Again the growth is in its essence a modified SK process, with strain relaxation as its main driving force. This method has produced important results, and is presently the strongest competitor to SK(H) and PQDs.[19]

All these methods have achieved good position control, allowing for the insertion in photonic structures.[19,20,21] Nevertheless, having site control without good optical properties is not a particularly useful feature as such. Self assembled SK dots showed even in the early stages extremely good optical properties, with "nearly" lifetime limited optical emission from single QDs reported.[22] SK are now obtained with extraordinary quality by several groups.[23] This is still a challenge in the field of site-controlled QDs and only recently some important breakthroughs on GaAs and InP substrates have been obtained in terms of optical properties.[19,24,25] Nevertheless, lifetime limited linewidths (generally a few microeV or less) in non resonant photoluminescence have not been obtained yet when a processing step precedes the epitaxial growth. Several reasons for that can be put forward, and while every group has developed its own understanding and recipes, the problem to overcome seems to be the impurity take-over from the processed interface to the epilayers. The advantage of PQD and of SKR in this contest is that it is often possible to grow a relatively thick buffer, therefore obtaining a reduction of the impurity transport to the QD layer, when compared to SKH approaches, where thinner buffers are grown.

Low spectral meandering with SK methods was obtained very recently in Toronto by chemical beam epitaxy (CBE) or in Wurzburg by molecular beam epitaxy (MBE). [19,25] In Fig. 3 (left) we show a representative summary of the photoluminescence full width at half maximum (FWHM) results reported in the literature, under non resonant pumping and at low temperature, categorised by types:[26] beam epitaxy (MBE/CBE) SK dots, site controlled SK based dots, MOVPE grown dots, etc. The latest results with PQDs (the record result is at the moment 18 microeV FWHM for the free exciton) are clearly well ahead in the site controlled category, compared to all other techniques.[24] An outcome even more striking since it is obtained by MOVPE,[27] and not by MBE, the standard technique for obtaining high quality QDs. Nevertheless both the SKR and SKH approaches are catching up quickly (~50 microeV), pushed by the needs for example of obtaining good spectral matching with a photonic cavity.

While narrow FWHM seem to be at reach in all the systems discussed, another important parameter actively investigated is the uniformity of the QD emission in a QD field. The motivation was already anticipated in our introduction: a uniform QD field will improve optoelectronic devices and their scalability, i.e. better lasers can be envisaged, together with a higher production yield in nanophotonic cavity based devices (micropillars, photonic crystal membranes, etc). Pyramidal QD ensembles show particularly high uniformity. In Fig 3 (right) we present a selection of literature data on the full width at half maximum (FWHM) of luminescence spectra versus the fundamental to excited state energy separation (S-P transitions) for Stranski-Krastanow (SK) and Pyramidal QDs.[28] Strikingly Pyramidal QDs have shown a dispersion as low as 1-3meV, maintaining a significant confinement, demonstrating a dramatic differentiation with other QD systems. Nevertheless SK dots uniformity has improved significantly, with and without site control, with FWHM of the ensemble photoluminescence as low as 15meV (with significant S-P separation) routinely obtained by several groups. It is worth stressing that in this case the QD confinement is an essential co-parameter, as SK dots can be annealed to bring extremely high uniformity, but with reduced confinement, limiting their applicability.

PQD are more likely to be a uniform system due to the specific nature of the growth process: since the growth happens layer by layer in a recess, the only uniformities one expects to act on the dots are those coming from the recess geometry itself. These are only originated by the lithographic process, so a "perfect" processing should deliver perfect uniformity (with as only residual statistical fractions of monolayer oscillations on the dot thickness, as in QW systems, to which at least a thick dot should be relatively insensitive). Moreover the growth process of PQDs tends to cure geometric non-uniformities, for two other reasons. The first one is rather an obvious one: the system tries to achieve a self limited profile at the bottom of the recess defined only by the fact that, at equilibrium, the pyramid centre (capillarity induced) grows at the same rate vertically to the lateral facets (determined by the precursors decomposition rate). So the actual profile the dot grows on is only determined by physical parameters (for example the dot base can be tuned from 20 to ~80 nm just changing the growth temperature)[5] and not by the original lithographic process, which will only determine the transient time before equilibrium is reached. The second less obvious aspect is that during the precursors decomposition, which determines the actual dot growth rate locally, the pyramidal recesses are in competition with each other. Larger recesses will have a higher capture cross section for precursors than smaller ones, i.e. larger recesses will actually grow faster than smaller ones as discussed in Ref. 4. Doing so they will fill themselves faster, and will be evolving into smaller recesses, till they will have the same dimensions as the surrounding ones. These last, being smaller at the beginning, grew at a lesser rate during buffer and barrier growths. The result is a sort of self equilibrant system, aiming towards high uniformity.[4]

Yet, other parameters need to be controlled in a QD system, if a real versatility is to be obtained. Overall most of them will enter the category of "wavefunction engineering". The actual meaning being the possibility of controlling the QD electronic states in the broadest sense, from its emission energy, the wavefunction symmetry properties, the excitonic binding energies, the single particle state characteristics (for example for long hole spin coherence a heavy hole like wavefunction is required), etc. All to be eventually extended to QD molecules and similar (e.g. dot in a dot, dot-wire-dot structures).

This is probably where PQD have the biggest potentialities, and some exciting perspectives are opening. Since the growth is pseudomorphic on the substrate, in many aspects it delivers the same control that QW growth allows. It enables for composition and thickness control naturally. It also allows for easy stacking of two nearly identical QDs one on top of each other, and to obtain different wavefunction coupling by only tuning the interdot distance; a topic which we address briefly with a small diversion before going back to single QD quantum states engineering concept. In Fig 4 there is an example of this: two nominally 0.5 nm $In_{0.25}Ga_{0.75}As$ dots have been grown one on top of each other (with 10, 2, 1 nm nominal interdot distance, respectively). The three samples have been excited in a regime where only two photoluminescence peaks where visible. In the first sample (10 nm distance) the two dots were optically excited independently, and were independently emitting at the free exciton wavelength (disperse around the average value a few meV), as the small statistic in Fig 4 shows. When the interdot distance was heavily reduced, the system emission red shifted significantly, and no more uncorrelated peaks could be found. This is a nice and easy compelling evidence that the system reacted to smaller interdot distances, but not *per se* a proof of molecular coupling. Understanding the exact nature of the achieved coupling requires more than this simple example. The interested reader could address himself to the recently appeared Ref. 29, where the full characterization of a specific PQD molecule coupling was performed. To our knowledge it is the first work demonstrating undoubtedly this achievement with site controlled structures.

It is important to observe that vertical coupling with all versions of SK methods offer an intrinsically lower level of control. Indeed it is possible to achieve QD coupling ("molecular" or tunnel)[30,31] but a significantly higher effort is needed to obtain the required structure, as the growth process of the upper dot will essentially be dramatically affected by the presence of the lower one, and several calibration runs are needed to obtain the right growth parameters. In the case of PQD the growth interaction is significantly smaller (the first dot affects the self limited profile of the second one only weakly) and stacking identical dots is a much easier task.

On the other hand the pseudomorphic growth turns out to be a limitation of the PQD wavelength tunability (with no controlled relaxation long wavelength QDs are a problem). Experimental findings limited the QD emission (InGaAs QD in GaAs barriers) in the infrared region to only slightly more than 1 micron emission wavelength. All together PQD emission has been demonstrated in a window not broader than the 700 -1100 nm interval. To do more than that a change in materials is needed, and relaxation processes must be avoided.

An interesting new development is the possibility of inserting nitrogen in a single QD structure. In itself it is not a simple exercise, since incorporating large quantities of nitrogen imposes in both the SK systems and PQD strong limitations in the growth conditions, typically low growth rate, temperature and As/III ratios, somehow at odds, especially for PQD, to "best" growth conditions found for other QD growths. Nevertheless nitrogen insertion helps extending the long wavelength emission for PQDs, and also helps reducing strain in the QD structures, which is one of the ingredients affecting the QD emission. Indeed in PQDs the (111)B growth is such that piezo fields are not negligible, and have to be considered in the dot design.[32]

Till now nitrogen incorporation has been limited to less than 1% in PQDs.[33,34] In big pitch pyramids growth conditions intentionally closer to "normal" dot growth have been utilized with the aim of maintaining good optical properties, while in small pitch pyramids the basic aim was to demonstrate longer QD wavelengths.[35]

This research is still in its early stages, and it is premature to predict how far this technique will be able to push the QD emission in the infrared. What we want to discuss here in deeper detail is an unexpected effect of nitrogen incorporation (especially in the limit of strong dilution).

Dimastrodonato et al. have shown[33] two important results. The first one is that beside the fact the incorporated nitrogen amount could be as low as to have negligible effects on the emission energy, (only a few meV red shift), the excitonic properties could change strongly. In particular nitrogen exposed samples showed an antibinding biexciton (assignment by power dependence and photon correlation spectroscopy)[33,36], while the without nitrogen counterparts showed a binding behaviour (see Fig 5, for all other details and discussion we refer the reader to Ref 33). This opens interesting perspectives, for example in the field of entangled photon generation. It has been shown[37] that in the case of zero biexciton binding energy, it would be possible to obtain polarization entangled photons for a biexciton-exciton cascade[38] even in the presence of a fine structure splitting (FSS, i.e. an asymmetry induced lift of the neutral exciton state degeneracy, which generally impedes entangled photon generation, as the demonstrated concepts rely on the state degeneracy). In Fig 6 we present a sketch of the argument included in Ref. 37, showing that with zero biexciton binding energy the two cascaded paths are equivalent, i.e. will generate entangled photons. It is obvious to extrapolate that by controlling the nitrogen exposure it should be possible to tune the QD excitonic state to a zero biexciton binding energy. Since the system allows for a few meV uniformity from dot to dot, it is easy to predict that on a single sample with an average zero biexciton binding energy, a significant percentage of the dots will show such behaviour, while others will slightly deviate. This should allow for easy addressability of the needed QD.

Even more significant is another surprising behaviour. Nitrogen incorporation also strongly reduced the asymmetry of the dots. In Fig. 7 a comparison of representative FSS measured with and without nitrogen is shown. Clearly a nearly zero FSS was found when the sample was exposed to nitrogen, suggesting between the others a reduction of geometric asymmetry and/or even compensating effect on built-in strain. Again this value comes with a statistical distribution, but on a single sample a majority of dots show a vanishing FSS. It should be stressed that even if PQD have been demonstrated as good entangled photon emitters, only carefully selected dots up to now have allowed that.[39] Generally PQD all show some residual FSS (even if significantly smaller than SK dots). The nitrogen incorporation approach seems to allow for a stronger dot to dot homogeneity in that respect.

**Open perspectives**

An interesting effect would be the opportunity of matching the previous speculation and the last result discussed. In principle, the current data and understanding tell us that by changing growth conditions, dot geometry and nitrogen flux exposure, it should be possible to obtain a sample with a large number of dots showing both zero biexciton binding energy and zero FSS. If so, this would open the possibility of generating photons not only polarization entangled (while preserving energy distinguishability) but also at the same energy, i.e. energy indistinguishable and entangled (sometimes referred to as biphoton emission). For the moment this is obtainable only with nonlinear processes in bulk crystals.[40] The possibility of reaching this with a single QD opens the way to entangled and indistinguishable photons on demand, eventually electrically pumped[23], with a scalable and relatively cheap production process.

A second interesting possibility for quantum information is the following. Electron spins (and holes) trapped in solid-state systems (quantum dots included) exhibit strong hyperfine interactions with the nuclear spin reservoir, which is normally fluctuating and randomly oriented. This represents a fundamental decoherence mechanism for the electron/hole spin: it is important then to develop strategies to suppress/limit it in semiconductor quantum dots, if the spin has to act as one of the fundamental bricks of a quantum gate or memory. However, this possibility requires a deeper understanding of the properties of the mesoscopic QD nuclear spin ensemble and of the possibilities of manipulating the nuclear spins. It has been shown in fact that charge complexes in the dot lead to a polarization of the underlying nuclear spin system (under circularly polarized excitation), opening the way to optically controlling the QD nuclear field. As cross relaxation between an optically pumped electron spin and the nuclear spins of the QD causes a dynamic nuclear polarization, the resulting Overhauser field acts as an effective magnetic field, usually parallel to the photo-created electron spin orientation, which on its part acts back on the electron spin dynamics.[41]

Specifically it was demonstrated that it is possible to obtain highly coherent hole spin states.[42] The general understanding of this occurrence is related to the fact that the hole envelop state is made of Bloch wave functions presenting mainly heavy hole (HH) character. A HH character guarantees that the wavefunction substantially goes to zero on the nuclear locations, reducing part of the hyperfine interactions. That said, only part of the hyperfine interaction can be suppressed. In general, all QDs present a mixed hole character between HH and light hole (LH).[42,43] The LH component will enhance the hyperfine interaction, producing an enhanced hole dephasing. One route to an engineered even longer coherent hole state is than that of designing a QD where the LH character is heavily suppressed.[44] One way is intuitively that of growing QDs presenting a nearly 2D character (big and flat dots). This is a relatively simple task for Pyramidal QDs, where the in-plane dimensions are defined by the growth temperature of the barrier material, and their thickness simply by growth time. Basically to grow a high aspect ratio dot it is a matter of high growth temperature, low growth rate and short growth time. It is indeed in the plans of our group to investigate this possibility.

We finally observe that with SK dots obviously this is significantly less easy to engineer. If the dots are site-controlled (both with the SKH and SKR methods) even lower control is left in the definition of dots properties, as site-control comes with a number of annexed conditions, which are needed to achieve the nucleation control. This is one of the reasons why no SK site-control approach has allowed for entangled photon emission to date, as the SK methods allows for that only on carefully selected dots and through post-growth, painstaking analysis.

There is nevertheless a last very much needed development, still fundamentally necessary for PQD, somehow already implemented with the other site-controlled methods.[45] Despite the fact that electroluminescence was demonstrated,[46] and so electrically pumped single photon emission,[47] these remain "single spot" results, and no full strategy has been yet developed to achieve full electrical control. This goes beyond the simple electroluminescence, but involves using an electric field to handle with an external force the QD population and few particles effects, and/or QD coupling. This has led to extremely valuable results with SK dots in general,[23,48] and has not been implemented yet with PQDs. In Fig 8 we depict a representative sketch of the structure as it might be obtained. Here we propose a back etched geometry, i.e. a geometry where the substrate has been selectively removed, to improve photon collection. The main challenge is the non-planarity of the recesses, which makes all processing significantly less tolerant than in planar geometry. Nevertheless we do not see any reason why such structure should not be implemented, and our group is actively engaged to pursue this.

**Conclusions**

In conclusion we discussed the *pyramidal* single quantum dot family as having high potentials for quantum information concepts. Strong improvements on uniformity and spectral purity have been recently reported, and a noteworthy versatility for wavefunction engineering already demonstrated. This opens strong perspectives, a few of which we discussed as bearing high potential, both in the field of entangled photons generation and that of obtaining long coherent electron/hole spin states. One essential step towards these goals is nevertheless that of designing a PQD structure which allows for electrical control. This involves overcoming the processing challenges of non-planar geometries.

**Acknowledgments**

under grants 05/IN.1/I25 and 08/RFP/MTR1659. We would like to thank E.P. O'Reilly for the enlightening discussions on quantum dot properties.

**Figure 1**

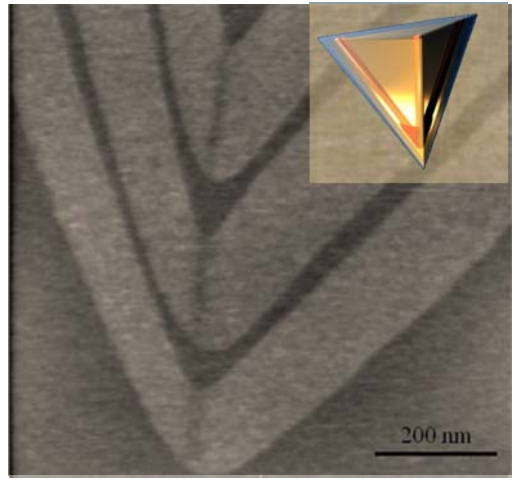

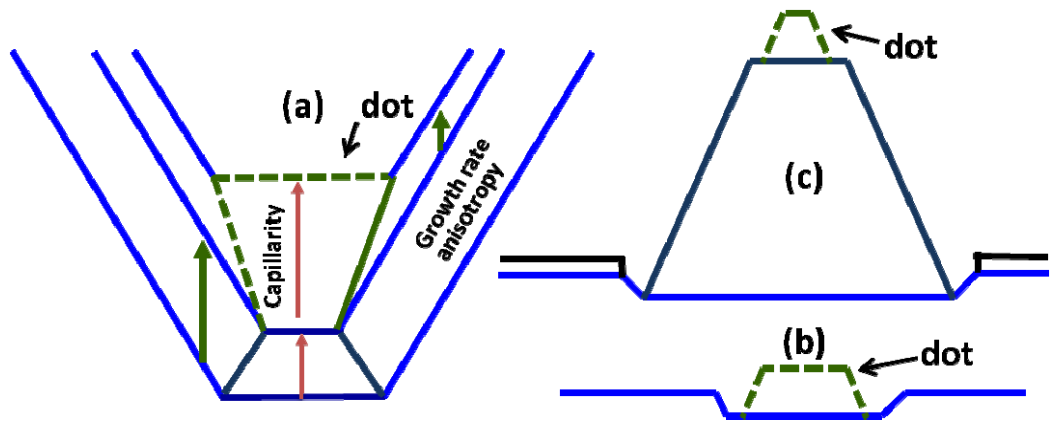

**Figure 3**

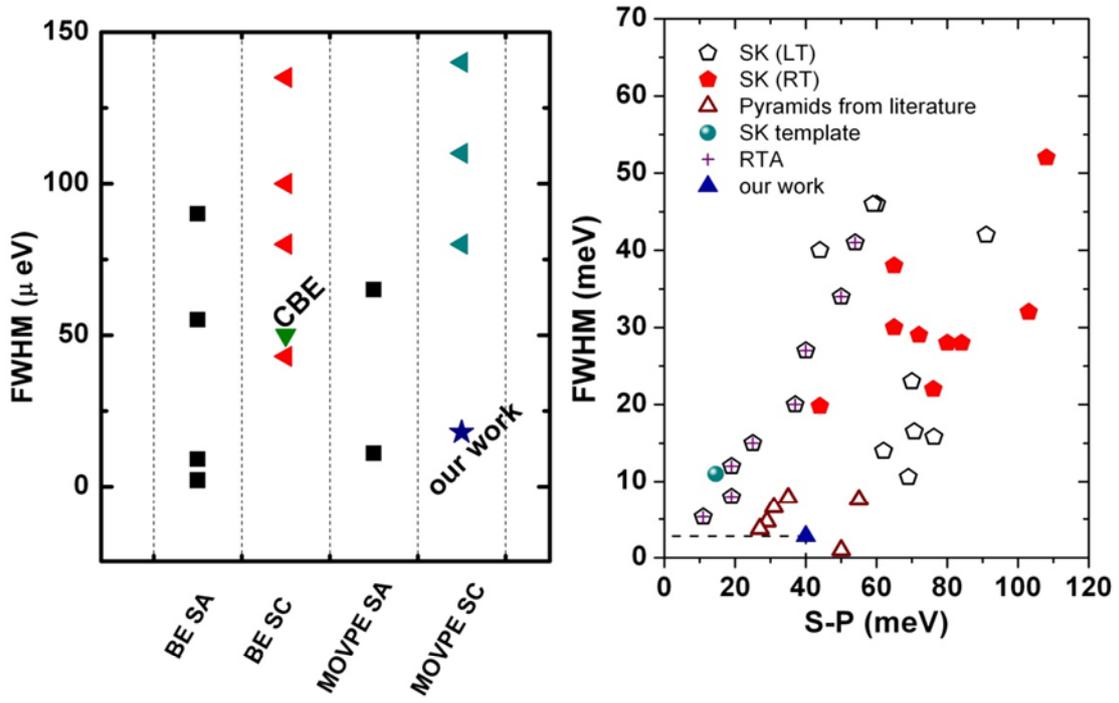

**Figure 4**

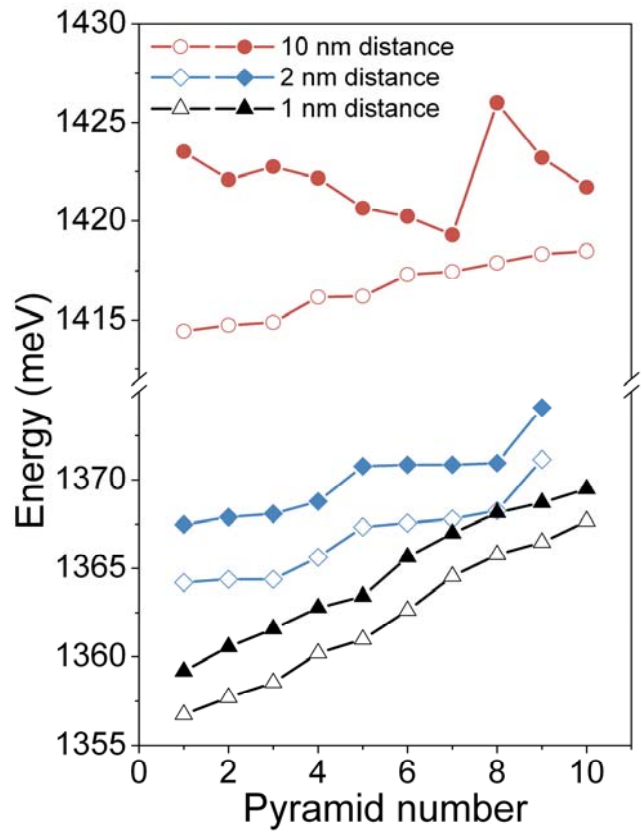

**Figure 5**

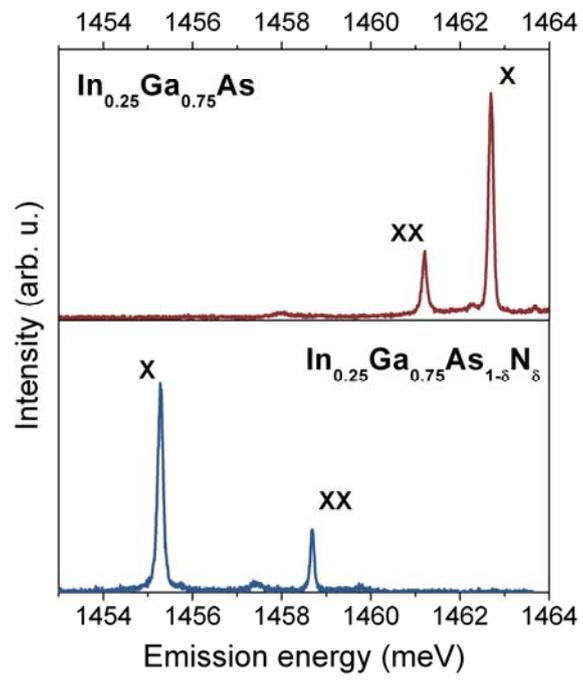

**Figure 6**

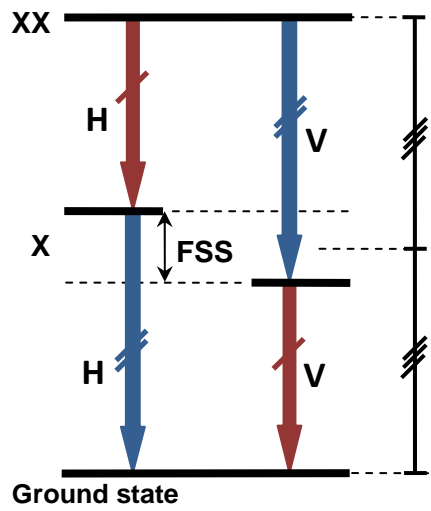

**Figure 7**

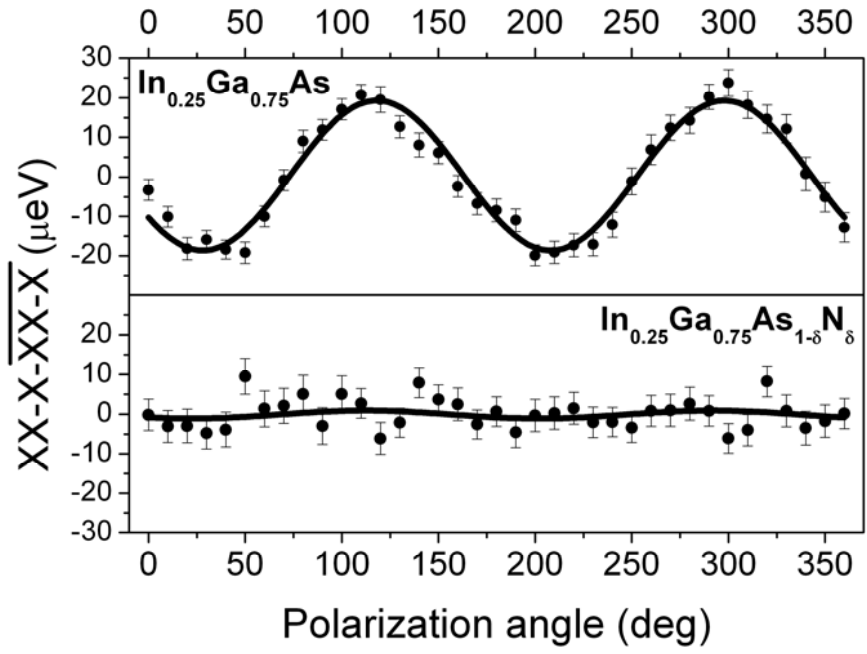

**Figure 8**

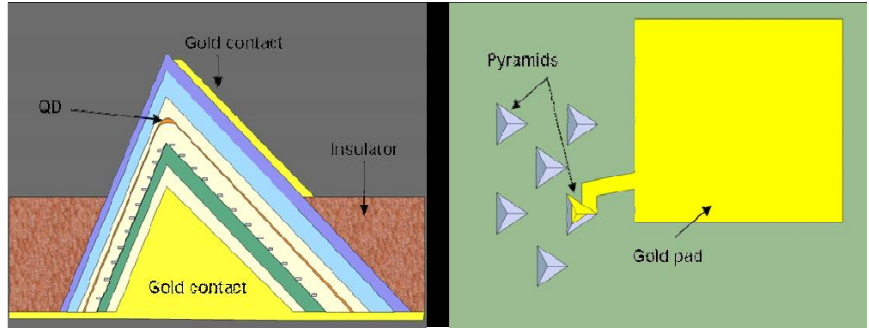

**Figure captions**

**Figure 1**: Cross-sectional atomic force microscopy image of a representative pyramidal system, with dot thickness intentionally designed to be easily contrasted by atomic force microscopy. The structure consists of a GaAs buffer and alternating layers of $Al_{0.3}Ga_{0.7}As$ (brighter) and pure GaAs (darker, the "quantum dots") markers. The contrast between different alloys results from the tip signal recording a different height due to a thin oxide layer grown on $Al_{0.3}Ga_{0.7}As$. Gallium segregation in the $Al_{0.3}Ga_{0.7}As$ layers is clearly visible along the vertical axis of the pyramid. (inset) Schematic representation of the QD structure and lateral quantum wires (at the three edges) and wells (along the three sidewalls).

**Figure 2**: Summary of the main differences of the three site-controlled QD growth methods discussed in the text. (a) Pyramidal dots are the only technique of those analyzed which grow pseudomorphically to the substrate; this thanks to growth rate anisotropy and capillarity. Growth rate anisotropy gives a sharp bottom by imposing a higher growth rate of the sidewalls, capillarity allows the dot formation by imposing a higher growth rate at the centre of the pyramidal recess. (b) SK in a hole patterns with shallow holes in a (100) GaAs substrate, and allows an SK process inside the shallow recess (growth can happen with or without the removal of the lithographic mask). (c) SK on a ridge/pyramid grows an SK dot on top of a 3D structure obtained after growth on a masked wafer is performed.

**Figure 3:** (left) Selection of representative literature data on the FWHM of luminescence spectra versus different type of growth protocols. Each column represents data in the corresponding category, as defined on the X axis label. SA: self assembled; SC: site controlled. BE: beam epitaxy (MBE and CBE). Our work (symbol: star) is by far the narrowest in the SC category. See Ref. 26. (right) Selection of literature data on the FWHM of luminescence spectra versus the S-P transitions energy separation for SK and Pyramidal QDs. SK(LT): low temperature luminescence data for SK dots; SK(RT): room temperature luminescence data for SK dots; Pyramids from literature: low temperature PL and cathodoluminescence data for pyramidal QDs; SK template: low temperature PL data for GaAs QDs grown inside etched SK dot template; RTA: low temperature PL data for SK dots after rapid thermal annealing. See Ref. 28.

**Figure 4:** Emission energy correlation of two most prominent transitions (solid and open symbols, respectively) from QDs with variable interdot distance: 10 nm, 2 nm and 1 nm.

**Figure 5:** Photoluminescence spectra of $In_{0.25}Ga_{0.75}As$ QD (top) and its counterpart exposed to nitrogen precursor during the growth (bottom).

**Figure 6:** Schematic of the realization of entangled photons generation through biexciton (XX) binding energy tuning to zero and time reordering of photons. H and V denote horizontally and vertically polarized photons, respectively.

**Figure 7:** (top) Demonstration of 19 microeV fine structure splitting in nitrogen-free $In_{0.25}Ga_{0.75}As$ QD; (bottom) No fine structure splitting has been resolved in all QDs exposed to nitrogen. The experimental values have been corrected for systematic errors by subtracting one waveform from the other. The results have been offset by their average.

**Figure 8:** Schematic representation of the planned 3D contacting scheme. One or more QDs can be present, and the different Pyramid layers can be doped or intrinsic, modulation doped etc depending on the application. In general an insulation layer (e.g. $SiO_2$, appropriate polymer) will be necessary to avoid shortcutting the top and bottom layers of the structure